\documentclass[twocolumn,showpacs,preprintnumbers,nofootinbib,amsmath,amssymb,aps,prd]{revtex4}
\usepackage{amsmath,amssymb}
\usepackage{amsfonts,amssymb,mathrsfs}
\usepackage[bookmarks=false,pdfstartview=FitH]{hyperref}

\def\be{\begin{equation}}
\def\ee{\end{equation}}
\def\nn{\nonumber}
\def\f{\frac}
\def\tf{\tfrac}
\def\sgn{{\rm sgn}}
\def\b{\bar}
\def\d{\dot}

\def\t{\tilde}

\def\dd{{\rm d}}

\def\de{\delta}

\def\om{\omega}

\def\vp{\varphi}
\def\De{\Delta}
\def\mR{\mathcal{R}}

\begin{document}

\pagestyle{plain}

\title{Ekpyrotic Loop Quantum Cosmology}

\author{Edward Wilson-Ewing} \email{wilson-ewing@phys.lsu.edu}

\affiliation{Department of Astronomy and Physics, Louisiana State University, Baton Rouge, 70803}

\begin{abstract}

We consider the ekpyrotic paradigm in the context of loop
quantum cosmology.  In loop quantum cosmology the classical
big-bang singularity is resolved due to quantum gravity
effects, and so the contracting ekpyrotic branch of the
universe and its later expanding phase are connected by a
smooth bounce.  Thus, it is possible to explicitly determine
the evolution of scalar perturbations, from the contracting
ekpyrotic phase through the bounce and to the post-bounce
expanding epoch.  The possibilities of having either one or
two scalar fields have been suggested for the ekpyrotic
universe, and both cases will be considered here.  In the
case of a single scalar field, the constant mode of the
curvature perturbations after the bounce is found to
have a blue spectrum.  On the other hand, for the two scalar
field ekpyrotic model where scale-invariant entropy perturbations
source additional terms in the curvature perturbations,
the power spectrum in the post-bounce expanding cosmology
is shown to be nearly scale-invariant and so agrees with
observations.

\end{abstract}


\maketitle

\section{Introduction}
\label{s.intro}

The ekpyrotic paradigm is an alternative to inflation where scale-invariant
perturbations are generated during a contracting pre-big-bang phase, and it
has been argued that these perturbations will travel through some transition
to an expanding Friedmann-Lema\^itre-Robertson-Walker (FLRW) space-time
and provide the seeds for structure formation \cite{Khoury:2001wf, Khoury:2001zk,
Lehners:2008vx}.  For the most up-to-date discussion on the observational status
of the ekpyrotic model, see \cite{Lehners:2013cka}.

While the ekpyrotic paradigm is motivated by string theory and sees the big-bang
singularity as the collision of two parallel branes, one of which is the
four-dimensional space-time we observe \cite{Khoury:2001bz}; in much of the work
studying the ekpyrotic universe an effective four-dimensional theory has been
used where there are typically one or two scalar fields representing the relevant
higher dimensional degrees of freedom that impact upon the dynamics of the
lower-dimensional brane.

Depending on the effective four-dimensional model used, the scalar perturbations
responsible for structure formation are generated by one of two methods.  In the
case where there is one scalar field with a particular negative exponential
potential, the growing mode of the gauge-invariant Bardeen potential becomes
scale-invariant during the contracting phase \cite{Khoury:2001zk}.  On the
other hand, when there are two (or more) scalar fields, then the entropy
perturbations of the scalar fields grow and also become scale-invariant during
the contracting phase.  Then, it is possible for these entropy fluctuations
to generate scale-invariant curvature perturbations \cite{Finelli:2002we,
DiMarco:2002eb, Lehners:2007ac, Koyama:2007mg}.

In both scenarios, it is necessary to show that the scale-invariance obtained
in the contracting branch survives to the expanding branch in order for
the predictions of the theory to match observations.  This is a difficult
task, since it is expected that quantum gravity effects will become important
at the transition point and the expectations one may have coming from
general relativity could be incorrect.

Because of this problem, there has been a lot attention paid to various
``matching conditions'' between the contracting and expanding branches
in order to determine the form the perturbations will have in the
expanding universe.  Depending on the specific prescription for the
matching conditions, different conclusions are reached.  For example,
for the case where there is a single scalar field, some matching
conditions argue that the perturbations in the expanding branch should
be scale-invariant \cite{Khoury:2001zk}, while others give a blue
spectrum \cite{Brandenberger:2001bs, Hwang:2001ga, Copeland:2006tn}.

In order to avoid the ambiguity present in the choice of matching conditions,
it is possible to use the presence of a dynamical attractor in the contracting
phase of the ekpyrotic universe to motivate the approximation that the
observationally relevant perturbations remain frozen during the transition
\cite{Creminelli:2004jg}.  If this is the case, then in the single scalar
field realization of the ekpyrotic universe the curvature perturbations
have a blue spectrum.  Of course, it is necessary to understand precisely
how the transition occurs in order to determine whether the approximations
are valid or not.

An approach which addresses this last problem is to model the transition
as a nonsingular bounce by using a ghost condensate \cite{Buchbinder:2007ad,
Creminelli:2007aq}, and then it can be shown that for the two-field ekpyrotic
paradigm the perturbations in the expanding branch are scale-invariant.
(The ghost condensate can cause a bounce and avoid the classical
big-bang singularity as it violates the null energy condition.)
However, here the equations of general relativity are used at all times
including the bounce, and so it is not clear whether some important
quantum gravity effects may be missed in the ghost condensate
ekpyrotic universes.

In this paper, we propose an alternative: rather than using a
ghost condensate to cause the bounce, we will instead consider
modifications to the gravitational dynamics due to quantum gravity
effects.  This will be done by working in the framework of loop
quantum cosmology, where the singularity is generically resolved
and replaced by a bounce \cite{Singh:2009mz}.  Using the effective
equations of loop quantum cosmology (LQC), it will be possible to
explicitly evolve the perturbations through the bounce and determine
the spectrum of the scalar perturbations in the expanding branch.
The homogeneous sector of the ekpyrotic universe has already
been studied in LQC, for the cases of the flat FLRW space-time
\cite{Bojowald:2004kt, Cailleteau:2008wu, dgms-unp} and the Bianchi I
model \cite{Cailleteau:2009fv}.  For general reviews of loop quantum
cosmology, see e.g.\ \cite{Bojowald:2008zzb, Ashtekar:2011ni,
Banerjee:2011qu}.

The outline of the paper is as follows: in Sec.\ \ref{s.setup}
we begin by building on the results of \cite{Bojowald:2004kt,
Cailleteau:2008wu} and show how the ekpyrotic universe can easily
be incorporated in LQC.  Then, we will describe the prescription
used in order to determine the evolution of the perturbations,
from the distant past of the contracting branch, through the
quantum-gravity-dominated bounce, to a sector of the expanding
branch where quantum gravity effects are negligible and the
equations of general relativity are once again valid.

This procedure will be carried out in Sec.\ \ref{s.single} for the case
where there is one scalar field, and in Sec.\ \ref{s.two} for the
two-field ekpyrotic model, where entropy perturbations source additional
terms in the curvature perturbations.  As we shall see, in the
expanding branch, the curvature perturbations in the single field model
have a blue spectrum, while in the two field model they are scale-invariant.
Thus, it is only the two scalar field model of ekpyrotic loop quantum
cosmology that is in agreement with observations.  We conclude with a
brief discussion in Sec.\ \ref{s.disc}.

\section{The Ekpyrotic Universe in Loop Quantum Cosmology}
\label{s.setup}

\subsection{Loop Quantum Cosmology}
\label{ss.lqc}

In loop quantum cosmology, cosmological space-times are quantized by
taking holonomies and areas to be the fundamental geometrical operators,
just as in loop quantum gravity.  Furthermore, the same techniques and
mathematical tools are used in order to define the Hamiltonian constraint
operator.  While several cosmologies have been studied in LQC, including
Bianchi and Gowdy space-times, here we will focus on the flat FLRW
cosmology, as it is this background that is relevant for the ekpyrotic
universe%
\footnote{Anisotropies and spatial curvature do not
play a significant role during the contracting
phase as the gravitational dynamics are dominated
by the ekpyrotic scalar field.  See e.g.\
\cite{Lehners:2008vx} for details.}.

The LQC dynamics of the flat FLRW space-time have been studied in some
detail now, and the most important result is the resolution of the
classical big-bang singularity.  Working in the Schr\"odinger picture,
states never become singular (assuming that they are initially
nonsingular) and undergo a bounce that bridges contracting and
expanding branches of the cosmology \cite{Ashtekar:2006wn,
Ashtekar:2007em}.

What is more is that states that are initially sharply peaked
remain sharply peaked throughout their evolution \cite{Ashtekar:2006wn}.
Thus, it is possible to speak of a ``mean geometry'' for such a
state at all times, including at the bounce point.  Furthermore,
the dynamics of the mean geometry is given by a relatively simple
set of modified Friedmann equations, called the LQC effective
equations.  For the flat FLRW space-time, the effective equations
are \cite{Taveras:2008ke}
\be \label{lqc1}
H^2 = \f{8 \pi G}{3} \rho \left( 1 - \f{\rho}{\rho_c} \right),
\ee
\be \label{lqc3}
\d{H} = -4 \pi G (\rho + P) \left( 1 - \f{2 \rho}{\rho_c} \right),
\ee
\be \label{lqc2}
\dot{\rho} + 3 H (\rho + P) = 0.
\ee
Here $H = \d{a}/a$ is the Hubble rate, $\rho$ and $P$ are respectively
the energy density and pressure of the matter field, and the dots
represent derivatives with respect to the proper time $t$.  The critical
energy density $\rho_c \sim \rho_{\rm Pl}$ encodes the modifications
to the Friedmann equations, and it is easy to see how its presence
causes a bounce to occur.

While these equations only hold for semi-classical states (i.e., states
that satisfy the scalar constraint and are initially sharply peaked),
in this paper we are interested in a universe which has a nice classical
limit, both before the bounce in the contracting branch, and in the
expanding branch after the bounce.  In order to have a nice classical
limit, the states should be sharply peaked and therefore we will
restrict our attention to semi-classical states for the remainder
of this paper.  So, it is sufficient to work with the effective equations
in this setting.

\subsection{The Ekpyrotic Universe}
\label{ss.ekpyrotic}

The potential for the scalar field in the ekpyrotic model is somewhat
complicated, but during the contracting phase of the universe, the
section of the potential that is relevant for that portion of the
gravitational dynamics has the form of a negative exponential and is
usually parametrized as
\be \label{potential1}
V(\vp) = -V_o e^{- \sqrt{16 \pi G / p} \, \vp},
\ee
with $0 < p \ll 1$.  One of the nice properties of this potential is
that it admits a solution where the scalar field mimics an ultrastiff
fluid with a constant equation of state $P = \om \rho$, with
$\om \gg 1$, and this solution is precisely the one of interest in
the ekpyrotic universe.

However, this is a result that comes from the classical Friedmann equations,
and no longer holds if one uses the LQC-corrected Friedmann equations
instead.  In order to allow a solution with the same $\om$ in LQC, a
slightly different potential must be used, namely \cite{Mielczarek:2008qw}
\be \label{potential2}
V(\vp) = \f{-V_o e^{- \sqrt{16 \pi G / p} \, \vp}}{\left(1 + \tf{3 p V_o}
{4 \rho_c (1-3p)} e^{- \sqrt{16 \pi G / p} \, \vp} \right)^2}.
\ee
Note that the potentials \eqref{potential1} and \eqref{potential2} agree
so long as quantum gravity effects remain negligible: the potentials
only differ once the bounce begins.

Using the usual relations $\rho = \d{\vp}^2/2 + V(\vp)$ and
$P = \d{\vp}^2/2 - V(\vp)$ together with the LQC effective equations,
it is easy to see that there exists a solution where
\be \label{ekpyrotic}
P = \om \rho, \qquad \om = \f{2}{3p} - 1.
\ee
This is the ekpyrotic solution that we shall consider here.
Since $p$ is very small, it follows that $\om \gg 1$ and so
this is an ultrastiff perfect fluid.

We will use this potential for the ekpyrotic scalar field, during
the collapse and the bounce.  This choice means that the matter
field is essentially a perfect fluid with a constant equation of
state, and this will significantly simplify the calculations.
Although we assume a transition to a radiation-dominated era soon
after the bounce (but after quantum gravity effects have become
small), we will only study the dynamics up to a time before this
transition.

It is worth pointing out that in many ekpyrotic models the
form of the potential changes just before the bounce, contrary to
what we assume here; we leave the possibility of allowing different
potentials for future work.  However, we do expect that the main
predictions obtained in this paper, including the value of the
scalar index, will not be affected by a change in the potential
during (or just before) the bounce.  We discuss this point further
in Sec.\ \ref{ss.pot}.

For the ekpyrotic solution \eqref{ekpyrotic} from the potential
\eqref{potential2}, the scale factor is given by
\be \label{a-lqc}
a(t) = \left( a_o t^2 + 1 \right)^{p/2}, \qquad
a_o = \f{8 \pi G \rho_c}{3 p^2},
\ee
where $\rho_c$ is the critical energy density.  The overall
scaling freedom in the scale factor has been fixed by setting
$a(0) = 1$.

It is easy to obtain the classical limit for either the contracting
or expanding branch, which correspond to $t \ll -1/\sqrt{a_o}$ and
$t \gg 1/\sqrt{a_o}$ respectively.  We will combine both cases
simply by considering $|t| \gg 1/\sqrt{a_o}$, in which case the
scale factor is given by
\be \label{a-cl}
a(t) = a_o^{p/2} |t|^p.
\ee
It is important to keep in mind the role played by the absolute values
around $t$ when derivatives are calculated in the classical limit.  For
example, the Hubble rate is given by $H = \d{a}/a = \sgn(t) p/|t| = p/t$.

Some other useful relations are those between the proper time $t$
and the conformal time $\eta$,
\be \label{deta}
\dd \eta = \f{\dd t}{a(t)}
\ee
given by
\be \label{t-eta}
|\eta| = \f{|t|^{1-p}}{a_o^{p/2}(1-p)},
\ee
and the scale factor as a function of conformal time is
\be
a(\eta) = a_o^{p/2(1-p)} \Big[ (1-p) |\eta| \Big]^{p/(1-p)},
\ee
again in the classical limit of $|t| \gg 1/\sqrt{a_o}$.

\subsection{With Two Scalar Fields}
\label{ss.two}

Many realisations of the ekpyrotic universe use two scalar fields and it is
possible to do this in LQC as well.  As before, it is desired to have the
scalar fields, together, mimic a perfect fluid with a constant equation
of state.

In order to do this, it is necessary to modify the potential \eqref{potential2}
in the following way.  If there are two scalar fields $\vp_1, \vp_2$, with
potentials $V_1(\vp_1)$ and $V_2(\vp_2)$, then taking the potentials
\be \label{potential3}
V_1(\vp_1) = \f{-V_o e^{- \sqrt{16 \pi G / q_1} \, \vp_1}}{\left(1 + \tf{3 p^2 V_o}
{4 q_1 \rho_c (1-3p)} e^{- \sqrt{16 \pi G / q_1} \, \vp_1} \right)^2},
\ee
\be
V_2(\vp_2) = \f{-\t{V}_o e^{- \sqrt{16 \pi G / q_2} \, \vp_2}}{\left(1 + \tf{3 p^2 \t{V}_o}
{4 q_2 \rho_c (1-3p)} e^{- \sqrt{16 \pi G / q_2} \, \vp_2} \right)^2},
\ee
with $0 < q_i \ll 1$ and
\be
p = q_1 + q_2,
\ee
there exists a solution to the LQC Friedmann equations with
\be \label{phi1}
\vp_1(t) = \sqrt \f{q_1}{4 \pi G} \ln \Bigg[ - \sqrt \f{2 \pi G V_o}{q_1(1-3p)}
\left( t + \sqrt{t^2 + \f{1}{a_o}} \right) \Bigg],
\ee
\be \label{phi2}
\vp_2(t) = \sqrt \f{q_2}{4 \pi G} \ln \Bigg[ - \sqrt \f{2 \pi G \t{V}_o}{q_2(1-3p)}
\left( t + \sqrt{t^2 + \f{1}{a_o}} \right) \Bigg],
\ee
where $a_o = 8 \pi G \rho_c / 3 p^2$, as before.  For this solution, the scale factor
is given by \eqref{a-lqc} and thus we see that this combination of the two scalar
fields mimic a perfect fluid with a constant equation of state \eqref{ekpyrotic},
just as in the previous section.

So, by having two scalar fields with the specific potentials given here,
it is possible to have exactly the same ekpyrotic solution for the background
gravitational degrees of freedom as in the single scalar field case.

Finally, from this discussion it is clear how to generalize this procedure
in order to allow additional scalar fields in the ekpyrotic LQC universe:
the potentials $V_i(\vp_i)$ should be of the same form as \eqref{potential3},
where $p$ is redefined as $p = \sum_i q_i$.

\subsection{Scalar Perturbations}
\label{ss.perts}

One of the main reasons to consider the ekpyrotic universe in the context of
LQC is that it is now known how to go beyond homogeneous space-times and study
linear perturbations on a flat FLRW background in LQC.  The two approaches
developed so far are lattice loop quantum cosmology \cite{WilsonEwing:2012bx},
where the separate universes framework \cite{Wands:2000dp, Rigopoulos:2003ak}
is adapted to the setting of loop quantum cosmology, and also a hybrid
approach where the perturbations are quantized \`a la Fock on the homogeneous
loop quantum cosmology background space-time \cite{FernandezMendez:2012vi,
Agullo:2012fc}.

In addition, effective equations for the linear perturbations ---including
the effects coming from the holonomy corrections that cause the bounce in
the homogeneous models--- have been obtained in two independent ways,
namely, derived from the Hamiltonian constraint in lattice loop quantum
cosmology \cite{WilsonEwing:2011es, WilsonEwing:2012bx} and obtained by
using the anomaly freedom algorithm \cite{Cailleteau:2011kr, Cailleteau:2011mi,
Cailleteau:2012fy}.  Using this Mukhanov-Sasaki-like LQC effective equation
on an ekpyrotic background, it will be possible to track the evolution of
the perturbations through the bounce and determine their precise form in
the expanding branch.

As mentioned above, for the homogeneous and isotropic FLRW cosmologies, it
has been shown that states that are initially sharply peaked remain sharply
peaked throughout their evolution determined by the Hamiltonian constraint
operator.  So, if one is interested in the mean geometry of the space-time
rather than quantum fluctuations, the effective equations for the
background variables describe their evolution to a high degree of accuracy.
However, the full quantum dynamics of linear perturbations have not yet been
studied in sufficient detail in order to determine whether, for sharply
peaked states, the effective equations again provide a good approximation
for the expectation values of operators corresponding to linear perturbations.

Nonetheless, it does not seem unreasonable to hope that, at least for linear
perturbations, the effective equations can be useful beyond the setting of
homogeneous space-times.  Therefore, in this paper we shall restrict our
attention to states that are sharply peaked, and use the effective equations
to determine the dynamics of the background geometry and also of the
propagation of linear perturbations on it.  This will allow us to determine
the mean geometry of the space-time, including perturbations, at all times.

In principle, it is possible to solve the effective equations of LQC for
all times and thus immediately determine the mean geometry of the entire
space-time.  However, in practice the effective equations for the perturbations
are difficult to solve exactly and so we will instead use the following
procedure:
\begin{enumerate}
\item The mean background geometry has already been determined, in
Sec.\ \ref{ss.ekpyrotic} if there is one scalar field, and in
Sec.\ \ref{ss.two} if there are two scalar fields.
\item We will start in the portion of the contracting branch
where quantum gravity effects are negligible.  The perturbations
visible today in the cosmic microwave background are of a
sufficiently long wavelength that as the universe contracts,
they will all exit the Hubble radius before any LQC effects
become important.

Therefore, the standard equations of classical cosmological
perturbation theory can be used in order determine the long
wavelength behaviour of the perturbations in the classical,
contracting branch once the modes exit the Hubble radius, but
before quantum gravity effects become important.
\item The LQC effective equations for the perturbations simplify
significantly in the long wavelength limit.  Therefore, it
will be a relatively straightforward task to determine the
solution to the LQC-corrected differential equations in the
long wavelength limit, up to the constant prefactors of the
two independent solutions to the differential equation.

This solution will hold at all times, before, during and after
the bounce, for long wavelengths.  Therefore, in the classical
limit before the bounce, the LQC solution must agree with the
classical solution obtained in the previous step, and this
condition can be used in order to determine the two unknown
prefactors in the LQC solution.
\item After this matching condition is imposed, the solution for
long wavelength modes is known at all times, including in the
expanding branch after the bounce, and so it is possible to check
whether the resulting perturbations are scale-invariant or not.
\end{enumerate}

These steps will allow us to determine the precise spectrum of
the curvature perturbations in the expanding branch of ekpyrotic
loop quantum cosmology.

Finally, before beginning this procedure, a few comments on the
validity and robustness of the LQC effective equations are in order.
Although there are some concerns about the validity of the effective
Friedmann equations \eqref{lqc1}, \eqref{lqc3} and \eqref{lqc2},
---especially in situations where quantum back-reaction would be
important \cite{Bojowald:2012xy}--- so far numerical studies have
shown that the effective equations provide an excellent approximation
to the dynamics of expectation values for sharply peaked states in
all homogeneous LQC space-times studied so far, including for certain
classes of cyclic models similar to the ekpyrotic universe
\cite{dgms-unp}.

The effective equations for the perturbations are obtained in the same
manner as the Friedmann effective equations.  However, as the full
LQC quantum dynamics of the perturbations have not yet been analyzed
in sufficient detail, it is not known whether effective equations give
as good of an approximation to the quantum dynamics of perturbations
as they do for homogeneous space-times.  Nonetheless, in the separate
universe paradigm long wavelength perturbations are viewed as a
patchwork of separate FLRW cosmologies \cite{WilsonEwing:2012bx,
Wands:2000dp, Rigopoulos:2003ak}; and since the effective equations
are valid for FLRW space-times, the separate universe approach suggests
that the effective equations for perturbations will likely be an
excellent approximation to the quantum theory, at least for long
wavelength modes (which are the modes of interest during the bounce
when LQC effects are important).

\section{The Single Scalar Field Model}
\label{s.single}

The dynamics of scalar perturbations in the contracting branch of the
single scalar field ekpyrotic universe have been studied in some detail,
and there is by now a wealth of literature on the subject.  In particular,
it is well known that the perturbations in the scalar field $\vp$
are scale-invariant \cite{Khoury:2001wf}; and although the comoving
curvature perturbations have a blue spectrum \cite{Lyth:2001pf},
the growing mode in the Bardeen potential is scale-invariant
\cite{Khoury:2001zk}.

In the first part of this section, for the sake of completeness and
also to establish notation, we will briefly review some of these
results.  Then, in the next two parts we will use the LQC
Mukhanov-Sasaki equation to evolve the perturbations through the
bounce and determine their spectrum in the expanding branch.

\subsection{The Contracting Branch}
\label{ss.before1}

We will use the gauge-invariant Mukhanov-Sasaki variable $v$ in order
to study the evolution of perturbations.  It is defined as
\be \label{def-v}
v = a \left( \de \vp^{\rm (g.i.)} + \f{\d{\vp}}{H} \Phi \right),
\ee
where $\de \vp^{\rm (g.i.)}$ is the gauge-invariant scalar field
perturbation and $\Phi$ is the gauge-invariant Bardeen potential.
For a review of cosmological perturbation theory, see e.g.\
\cite{Mukhanov, Mukhanov:1990me}.

When there is only a single matter field, as is the case here,
the equations of motion for $v$ are
\be \label{ms-cl}
v'' - \nabla^2 v - \f{z''}{z}v = 0,
\ee
where primes denote derivatives with respect to the conformal
time, and
\be \label{def-z}
z = \f{a \d{\vp}}{H} = \f{a}{\sqrt{4\pi G p}},
\ee
where the second equality holds due to the definitions of $\rho$
and $P$ for a scalar field together with \eqref{ekpyrotic}.

Therefore, in Fourier space this equation becomes
\be
v_k'' + \left(k^2 - \f{p(2p-1)}{(p-1)^2 \eta^2} \right) v_k = 0,
\ee
and the solution is
\be
v_k = \sqrt \f{\pi \hbar}{4} \sqrt{-\eta} \, H_n^{(1)}(-k \eta);
\quad n = \f{1-3p}{2(1-p)},
\ee
where $H_n^{(1)}(x)$ is the Hankel function.  The prefactor of
$\sqrt{\pi \hbar / 4}$ ensures that initially (at time $\eta \to -\infty$),
$v_k$ are quantum vacuum fluctuations.

The comoving curvature perturbation $\mR$ is related to the
Mukhanov-Sasaki variable by
\be
\mR_k = \f{v_k}{z} = \f{\pi \sqrt{G\hbar p} \, (-\eta)^n}
{[\sqrt{a_o} (1-p)]^{p/(1-p)}} \, H_n^{(1)}(-k \eta).
\ee
When the modes exit the Hubble radius, $-k\eta \ll 1$ and then
the asymptotic form of the Hankel function can be used.  This
shows that for modes that have exited the Hubble radius,
\begin{align} \label{R-asym}
\mR_k =& \f{\pi \sqrt{G\hbar p}}{2^n \Gamma(n+1)
[\sqrt{a_o} (1-p)]^{p/(1-p)}} k^n (-\eta)^{2n} \nn \\ &
- \f{i \, 2^n \Gamma(n) \sqrt{G\hbar p}}
{[\sqrt{a_o} (1-p)]^{p/(1-p)}} k^{-n}.
\end{align}
Since $0 < p \ll 1$, we have $n \approx \tf{1}{2} - p$, and thus
both modes of $\mR$ have a blue spectrum.

From $\mR$ we can calculate the Bardeen potential $\Phi$ via the
relation \cite{Mukhanov}
\be \label{bardeen}
-k^2 \Phi_k = \f{4 \pi G a \d{\vp}^2}{H} \mR_k' = \f{\mR_k'}{(1-p)\eta}.
\ee
Using $\dd [x^\ell H_\ell^{(1)}(x)]/\dd x = x^\ell H_{\ell-1}^{(1)}(x)$,
we find that
\be \label{bardeen-before}
\Phi_k = \f{-\pi \sqrt{G\hbar p} \, (-\eta)^{n-1}}
{k [a_o^{p/2} (1-p)]^{1/(1-p)}} \, H_{n-1}^{(1)}(-k \eta).
\ee
Again, for modes that are outside the Hubble radius, the asymptotics
of the Hankel function can be used in order to determine the $k$-dependence
of the perturbations,
\begin{align}
\Phi_k =& \f{\pi \sqrt{G\hbar p}}{2^{n-1} \Gamma(n)
[a_o^{p/2} (1-p)]^{1/(1-p)}} k^{-(2-n)} (-\eta)^{2n-2} \nn \\ &
- \f{i \, 2^{n-1} \Gamma(n-1) \sqrt{G\hbar p}}
{[a_o^{p/2} (1-p)]^{1/(1-p)}} k^{-n}.
\end{align}
The first mode is growing as the space-time contracts and is almost scale-invariant
with a small red tilt, while the second mode is constant and proportional to
$k^{-1/2+p}$.  Recalling that for small $p$, $n \approx \tf{1}{2} - p$, the
power spectrum of $\Phi$ is easily determined.  Keeping only the growing
mode, which is also the dominant term for long wavelengths,
\be
\De_\Phi^2(k) = \f{k^3}{2 \pi^2} |\Phi_k|^2 \sim \f{k^{-2p}}{(-\eta)^{1+2p}},
\ee
and it is clear that the Bardeen potential is scale-invariant with a
slight red tilt due to the $-2p$ in the exponent.

Now the question is whether this mode survives the bounce and if it
is observationally relevant in the expanding branch of the ekpyrotic
universe.  We expect the constant modes in $\mR$ and $\Phi$ to agree,
so if the scale-invariant mode of $\Phi$ is indeed relevant, there
should appear a new term in the curvature perturbation $\mR$ as
well.  In order to determine the spectrum of the observationally
relevant modes in the expanding branch, we will use the effective
equations of LQC to evolve the perturbations through the bounce.

\subsection{The Bounce}
\label{ss.bounce1}

The LQC effective equation for the Mukhanov-Sasaki variable defined in
\eqref{def-v} is \cite{Cailleteau:2011kr, Cailleteau:2011mi}
\be \label{lqc-ms}
v'' - \left(1 - \f{2 \rho}{\rho_c}\right) \nabla^2 v - \f{z''}{z} v = 0,
\ee
where $z$ is given by the first equality appearing in \eqref{def-z}.
Note however that the second equality no longer holds as the Friedmann
equations relating $H$ and $\d{\vp}$ are modified in LQC.  Thus,
once corrections from LQC are included,
\be
z(t) = \f{1}{4\pi G} \sqrt \f{3 p}{2 \rho_c} \f{a(t)^{(p+1)/p}}{t}.
\ee

For long wavelength modes, the second term in \eqref{lqc-ms} is
negligible and then the solution to the differential equation for
each Fourier mode is
\be
v_k = A_1 z(\eta) + A_2 z(\eta) \int^\eta \f{\dd \b\eta}{z(\b\eta)^2},
\ee
where $A_1$ and $A_2$ are constants that must be fixed by matching
this solution with the classical solution in a regime when both
solutions are valid.

Recalling the definition of the comoving curvature perturbation
variable $\mR$, and using \eqref{deta},
\begin{align} \label{R-lqc1}
\mR_k &= A_1 + A_2 \int^t \f{\dd \b t}{a(\b t) \, z(\b t)^2} \\ &
= A_1 + A_2 \cdot \f{3 a_o p^3}{2 \rho_c}\, t \Bigg[ {}_2F_1 \left(\tf{1}{2},
\tf{3 p}{2}, \tf{3}{2}, -a_o t^2\right) \nn \\ & \qquad
\label{R-lqc2}
- {}_2F_1 \left(\tf{1}{2}, 1 + \tf{3 p}{2}, \tf{3}{2}, -a_o t^2\right) \Bigg]
+ \alpha A_2,
\end{align}
where ${}_2F_1$ are hypergeometric functions and $\alpha$ is a constant of
integration.  A nonzero $\alpha$ simply redefines the constant mode by
$A_1 \to A_1 + \alpha A_2$, and we will take advantage of this freedom
in order to simplify the matching that determines $A_1$ and $A_2$.  Of
course, the physics is independent of $\alpha$ in the sense that the matching
uniquely determines $A_1 + \alpha A_2$, but a particular choice of $\alpha$
will shorten the calculations.

The solution \eqref{R-lqc2} can be trusted for any mode whose conformal
wavelength is larger than $\sqrt{|z/z''|}$.  Since in the classical limit
$z/z'' \sim r_H^2$ (with $r_H$ being the conformal Hubble radius), this
solution holds for modes outside of the Hubble radius during the classical
contracting phase.

Therefore, this solution can be matched to \eqref{R-asym} by taking
the classical limit of $t \ll -1/\sqrt{a_o}$, in which case
\begin{align}
\mR_k =& A_1 + A_2 \Bigg[ \f{3 p^3 a_o^{1-3p/2} e^{-i \cdot 3 \pi p}
(-t)^{1-3p}} {2 (1-3p) \rho_c} + \alpha \nn \\ &
+ \f{3 p^3 \sqrt{a_o \pi}}{4 \rho_c} \left(
\f{\Gamma\left(\tf{1+3p}{2}\right)}{\Gamma\left(\tf{3p}{2}\right)}
- \f{\Gamma\left(-\tf{1-3p}{2}\right)}{\Gamma\left(1+\tf{3p}{2}\right)}
\right) \Bigg].
\end{align}
This expression can be simplified by setting
\be
\alpha = - \f{3 p^3 \sqrt{a_o \pi}}{4 \rho_c} \left(
\f{\Gamma\left(\tf{1+3p}{2}\right)}{\Gamma\left(\tf{3p}{2}\right)}
- \f{\Gamma\left(-\tf{1-3p}{2}\right)}{\Gamma\left(1+\tf{3p}{2}\right)}
\right),
\ee
and since in the classical limit the relation \eqref{t-eta} can be trusted,
we find that the matching of this solution with the classical one in the
regime where they both hold implies that
\be
A_1 = - \f{i \, 2^n \Gamma(n) \sqrt{G\hbar p}}
{a_o^{p/2}} k^{-n},
\ee
\be
A_2 = \f{2^{1-n} \pi \sqrt{G\hbar p} \, \rho_c}
{3 \, \Gamma(n+1) p^3 a_o^{(2-p)/2}}  k^n.
\ee
The expressions given here hold for small $p$.  It is a straightforward
calculation to determine the exact numerical factors to all orders of
$p$, but the precise form of the prefactors will not be important here.
Indeed, the most important point here is that $A_1 \sim k^{-n}$ and
$A_2 \sim k^n$.

\subsection{The Expanding Branch}
\label{ss.after1}

In the expanding branch after quantum gravity effects become negligible,
the standard equations of motion for linear cosmological perturbations
coming from general relativity are valid, so the differential equation
\eqref{ms-cl} can be used and therefore the solution for $\mR_k$, in
the expanding branch, is
\be \label{R-after}
\mR_k = \eta^n \Big[ B_1 J_n(k \eta) + B_2 Y_n(k \eta) \Big],
\ee
where the prefactors must be determined by matching this solution in the
long wavelength limit with the LQC solution in the late time limit.

The long wavelength limit of \eqref{R-after} is simply given by
\be \label{R-after2}
\mR_k = \f{B_1}{2^n \Gamma(n+1)} k^n \eta^{2n}
- \f{B_2 2^n \Gamma(n)}{\pi} k^{-n},
\ee
and the late time solution for the LQC solution \eqref{R-lqc2} is
\be
\mR_k = A_1 + A_2 \Bigg[ \f{3 p^3 a_o^{1-3p/2} t^{1-3p}}
{2 (1-3p) \rho_c} - 2 \alpha \Bigg].
\ee
Thus, by matching the prefactors, we find that we must have
\be
B_1 = \f{\sqrt{\pi \hbar}}{2}, \qquad
B_2 = i\f{\sqrt{\pi \hbar}}{2} \left( 1 - \f{2 \alpha A_2}{A_1} \right).
\ee
Therefore, the first mode of \eqref{R-after2} is proportional to
$k^n$ while the second has two contributions, one proportional to
$k^n$ as well and the other proportional to $k^{-n}$.

As in the ekpyrotic scenario $n \approx \tf{1}{2} - p$ and $p$ is
small, it is clear that neither of the modes of the curvature
variable $\mR$ contain a scale-invariant term.  While this is
not a surprising result (no term in the curvature perturbations
was scale-invariant in the contracting branch), it does indicate
that the constant mode of the Bardeen potential should also
contain two contributions, proportional to $k^n$ and $k^{-n}$,
neither being scale-invariant.

This expectation can be checked explicitly by using the relation
\eqref{bardeen} in order to relate the curvature perturbation in
\eqref{R-after} to the Bardeen potential.  Dropping the numerical
part of the overall prefactor [which is the same as in
Eq.\ \eqref{bardeen-before}], this gives
\be
\Phi_k \sim \f{\eta^{n-1}}{k} \Bigg[ H_{n-1}^{(1)}(k \eta)
- \f{2 \alpha A_2}{A_1} Y_{n-1}(k \eta) \Bigg].
\ee
In the long wavelength limit, the Bessel functions can be expanded
giving
\be
\Phi_k \sim k^{n-2} \eta^{2n-2} + k^{-n} + k^n,
\ee
where we have only kept the $k$ and $\eta$ dependence of each of
the terms.  (Recall that $\alpha A_2 / A_1 \sim k^{2n}$.)

Since $n \approx \tf{1}{2} - p$, we see that only the first term
is scale-invariant.  However, since this mode decays rapidly as
the universe expands, it is not relevant for structure formation.
Rather, the dominant contribution to the power spectrum of the
Bardeen potential at late times is
\be
\Delta_\Phi^2(k) \sim k^{3-2n} \sim k^{2+2p},
\ee
giving a blue spectrum with a scalar index of $n_s = 3 + 2p$, which
is in perfect agreement with the expectations coming from the form of
$\mR$ after the bounce.

It is worth pointing out that the $k^n$ contribution
appearing in the constant mode of $\Phi$ (and also $\mR$)
comes from the scale-invariant growing mode of $\Phi$ in
the contracting branch.  The reason that this term is not
scale-invariant in the constant mode is that it is, in
effect, multiplied by $k^2$, just as suggested by the matching
conditions used in \cite{Brandenberger:2001bs}.

Thus, the results obtained here by evolving the perturbations
through the bounce and in \cite{Brandenberger:2001bs} by
using the constant energy density matching conditions given
in \cite{Hwang:1991an, Deruelle:1995kd} are in agreement,
at least for predictions of the scalar index $n_s$.

Finally, it is also worth pointing out that in another bouncing
cosmology, the matter bounce scenario, the scalar index $n_s$
obtained from LQC effective equations and the constant energy
density matching conditions agree once again, even though
there is a nontrivial transfer from the growing mode in the
contracting phase to the constant mode in the expanding phase
\cite{Finelli:2001sr, WilsonEwing:2012pu}.  (Note however that
the predicted amplitudes of the perturbations do not agree.)
It is possible that this correspondence may be true in general,
and that the constant energy density matching conditions might
capture some of the physics occurring during the bounce in LQC.

\section{The Two Scalar Field Model}
\label{s.two}

In the two scalar field model of the ekpyrotic universe, the entropic
perturbations become scale-invariant in the contracting branch
\cite{Finelli:2002we, DiMarco:2002eb, Lehners:2007ac, Koyama:2007mg,
Buchbinder:2007ad, Creminelli:2007aq}.  Then, the (scale-invariant)
entropy perturbations can source (scale-invariant) curvature perturbations.

We begin this section by reviewing the dynamics of entropy perturbations
and recall results from \cite{Finelli:2002we, DiMarco:2002eb, Lehners:2007ac,
Koyama:2007mg, Buchbinder:2007ad, Creminelli:2007aq} that show that, for
the ekpyrotic solution, entropic perturbations are scale-invariant for
modes outside the Hubble radius.  Then, we will a consider a specific case
in order to show how the entropic perturbations can generate curvature
perturbations, and evolve these curvature perturbations through the LQC
bounce.

\subsection{Entropy Perturbations in the Contracting~Branch}
\label{ss.before2}

As there are two scalar fields $\vp_1$ and $\vp_2$, there are also two
independent perturbations $\de \vp_1$ and $\de \vp_2$.  It is of course
possible to study the perturbations of the matter degrees of freedom
using these variables, but it is more convenient to rewrite these
perturbations in terms of adiabatic and entropic perturbations.

It will be useful to briefly review some basic information regarding
entropy perturbations; for more details see e.g.\ \cite{Gordon:2000hv,
DiMarco:2002eb}.  For the case where there are two scalar fields, a
simple way to rewrite the perturbations $\de \vp_1$ and $\de \vp_2$ in
terms of adiabatic and entropic perturbations is available if the
trajectory of the scalar fields through their configuration space
$(\vp_1(t),\vp_2(t))$ is known.  The key ingredient is the angle
$\theta$ describing the direction of the motion of the scalar fields
in their configuration space,
\be
\cos\theta = \f{\d \vp_1}{\sqrt{\d \vp_1^2 + \d \vp_2^2}}, \qquad
\sin\theta = \f{\d \vp_2}{\sqrt{\d \vp_1^2 + \d \vp_2^2}}.
\ee
For the ekpyrotic solution given in \eqref{phi1} and \eqref{phi2},
\be
\theta = \arctan \sqrt \f{q_2}{q_1}.
\ee

Then, it is possible to use the angle $\theta$ in order to split the
perturbations $(\de \vp_1, \de \vp_2)$ into the part tangential to the
configuration space trajectory of the scalar fields, giving the adiabatic
perturbation
\be
\de \sigma = \cos \theta \, \de \vp_1 + \sin \theta \, \de \vp_2;
\ee
while the gauge-invariant entropic perturbation is orthogonal to the
configuration space trajectory of the scalar fields,
\be
\de s = - \sin \theta \, \de \vp_1 + \cos \theta \, \de \vp_2.
\ee
The adiabatic perturbations will have the same behaviour as $\de \vp$
in the single field case \cite{Gordon:2000hv} and so the adiabatic
perturbations alone would give the same results as the model studied
in Sec.\ \ref{s.single}.

Therefore, it is more interesting to focus on the entropy perturbations,
which we shall show contain a growing scale-invariant mode in the
contracting space-time.  Since entropic modes appear as a source in the
equations of motion for the curvature perturbation $\mR$, if
they are scale-invariant they can generate scale-invariant
curvature perturbations.

In order to see that the growing mode of the entropic perturbations is
scale-invariant, we start with the equation of motion for the entropic
perturbations in Fourier space \cite{Gordon:2000hv}
\be \label{ent1}
\ddot{\de s}_k + 3 H \d{ \de s}_k + \left( \f{k^2}{a^2} + V_{ss}
+ 3 \d\theta^2 \right) \de s_k =
\f{k^2 \d\theta}{2 \pi G a^2 \d\sigma} \Phi_k,
\ee
where we have defined
\be
V_{ss} = \cos^2 \theta \f{\dd^2 V_1(\vp_1)}{\dd \vp_1^2} + 
\sin^2 \theta \f{\dd^2 V_2(\vp_2)}{\dd \vp_2^2}.
\ee
(Note that there is an additional term appearing in $V_{ss}$ if there are
any interactions between $\vp_1$ and $\vp_2$.)

For the ekpyrotic solution given by \eqref{a-lqc}, \eqref{phi1} and
\eqref{phi2}, $\d\theta = 0$ and
\begin{align}
V_{ss} &= \f{-2 (1 - 3p)}{t^2 + 1/a_o} \left( 1 - \f{3/2}{a_o t^2 + 1} \right)
\nn \\ & \label{vss}
\approx -\f{2}{t^2} \left( 1 - 3p \right),
\end{align}
where the second equality provides an excellent
approximation to $V_{ss}$ so long as quantum gravity effects are
negligible.

Using $\d\theta = 0$ and Eq.\ \eqref{vss}, the differential
equation \eqref{ent1} ---rewritten for $(a \, \de s_k)$ in conformal time---
becomes
\be
(a \, \de s_k)'' + \left( k^2 - \f{1}{\eta^2} \left[2 - 3 p
+ p^2 \right] \right) a \, \de s_k = 0.
\ee
The solution to this differential equation is similar to the one for
the Mukhanov-Sasaki equation,
\be
a \, \de s_k = \f{\sqrt {\pi \hbar}}{2} \sqrt{-\eta} H_{\t{n}}^{(1)}(-k \eta),
\ee
with the difference that now
\be
\t{n} = \sqrt{\f{9}{4} - 3p + p^2} \approx \f{3}{2} - p,
\ee
where the second equality holds for small $p$.
As before, the overall prefactor has been chosen so that at early times
$\eta \to -\infty$ one obtains quantum vacuum fluctuations.

Recalling that $a(\eta) \sim (-\eta)^{p/(1-p)}$ in the contracting branch,
and using the asymptotic expansion of the Hankel function, the long wavelength
limit of $\de s_k$ is given by
\be
\de s_k = \sqrt \f{\pi \hbar}{2} a_o^{-p/2} \bigg[ \f{k^{3/2-p}
(-\eta)^{2-2p}}{3} - i \f{k^{-3/2+p}}{\sqrt\pi(-\eta)} \bigg],
\ee
where we have dropped terms of the order $p$ in the prefactors, and of
the order $p^2$ in the exponents.  This shows that the growing mode is
almost scale-invariant, with a slight blue tilt.

Here there is a blue tilt because of the choice of the potentials we have
worked with.  It is easy to choose slightly different potentials and get
a slightly different tilt.  In particular, it is relatively easy to
obtain a red tilt that agrees with observations \cite{Lehners:2007ac,
Buchbinder:2007ad, Fonseca:2011qi}.

\subsection{Generating Curvature Perturbations}
\label{ss.curv}

The dynamics of the curvature perturbation $\mR$ is affected by the presence
of entropic perturbations, and so it is possible for a scale-invariant entropy
perturbation to generate scale-invariant curvature perturbations.

If the Bardeen potential and the entropic perturbations are known, then the
evolution of the curvature perturbation is given by \cite{Gordon:2000hv}
\be \label{mRdot}
\d\mR_k = \f{k^2}{a^2} \cdot \f{H}{\d{H}} \Phi_k + \f{2 H \d\theta}
{\sqrt{\d\vp_1^2 + \d\vp_2^2}} \de s_k,
\ee
and in the long wavelength limit, the first term is negligible.  Note however
that in the solution we have been considering so far, $\d\theta = 0$ and so
in this case the entropic perturbations cannot source the curvature perturbations.

However, by modifying the potential so that one of the scalar fields suddenly
changes directions in their configuration space, it is possible to generate
new terms in the curvature perturbation.  Here we will review one such scenario
that has been proposed in order to show how this is possible.  We shall
merely outline the procedure studied in detail in \cite{Lehners:2007ac,
Buchbinder:2007ad}.  For alternative scenarios, see \cite{Koyama:2007ag,
Battefeld:2007st}.

So far we have been working with negative exponential potentials (plus some
minor modifications to simplify LQC calculations).  Let us now assume that the
potential $V_2(\vp_2)$ has the form where part of it is accurately described
by \eqref{potential3}, but then the potential rises sharply and flattens out.
(We assume that $V_1(\vp_1)$ remains unchanged.)  If the form of the potential
is chosen carefully, then it is possible to obtain a solution $\d\vp_2 = 0$
after $\vp_2$ climbs the potential, and this solution will be conserved in
time if the potential $V_2(\vp_2)$ becomes flat after its sharp rise.

In this scenario, the new value of $\theta = 0$ shows that there must have
been a $\d\theta \neq 0$ at some point.  If we assume that the potential rises
very sharply, it is possible to approximate the change in $\theta$ by a
delta function \cite{Buchbinder:2007ad},
\be
\d\theta = - \left( \arctan \sqrt \f{q_2}{q_1} \right) \delta(t-t_i),
\ee
where $t_i$ is the time when $\vp_2$ hits the sudden rise in the potential.
Of course, as $\vp_2$ starts to rise in the potential, the dynamics of the
background will be affected, so $H$, $\d\vp_1^2 + \d\vp_2^2$ and $\de s$
will change.  However, if we assume that the change in $\theta$ is
instantaneous, we can approximate these quantities by their values
just before $\vp_2$ starts to go up the sharp rise in the potential%
\footnote{At first it seems as though such an
approximation must fail for both $\d\vp_1^2
+ \d\vp_2^2$ and $\de s$, as they each
contain contributions from $\d\vp_2$ which
changes rapidly as $\vp_2$ climbs the potential.
However, it can be shown that the first term
must vary continuously and $\de s$ changes
by at most a factor of order unity (see
\cite{Buchbinder:2007ad} for details),
and so this approximation is valid.  Note
however that this approximation fails in
the treatment of non-Gaussianities
\cite{Lehners:2007wc}.}.
As an aside, note that while a sharp change in $\d\theta$ simplifies the
calculation, a more gradual transition in the value of $\theta$ results
in a larger amplitude of the terms in $\mR_k$ generated by the entropy
perturbations \cite{Lehners:2007wc}.

In this limit, it is easy to integrate \eqref{mRdot}, giving
\be
\mR_k(t) = \mathring{\mR}_k(t) - 8 \pi G \sqrt{p}
\left(\arctan \sqrt \f{q_2}{q_1}\right) t_i \, \de s_k(t_i),
\ee
where $\mathring{\mR}_k$ represents the terms that were already present in
the curvature perturbation variable.  Recall that the curvature
perturbation, in a contracting ekpyrotic cosmology, has a constant
mode as well as a decaying mode as seen in Sec.\ \ref{ss.before1}.
Since the new terms generated by the entropic modes are evaluated
at the time $t_i$, these new terms do not evolve after this time and so
contribute to the constant mode.

Therefore, the curvature perturbation in this scenario is schematically
given by%
\footnote{Note that while the time dependence of
the first, decaying mode may change when the
field $\vp_2$ climbs the potential, this mode
will continue to be time-dependent and its
$k$-dependence will not change.}
\begin{align} \label{mRbefore}
\mR_k \sim & \,\, k^{\tf{1}{2}-p} (-\eta)^{1-2p} + k^{-\tf{1}{2}+p}
+ k^{\tf{3}{2}-p} + k^{-\tf{3}{2}+p},
\end{align}
where the first two terms come from $\mathring{\mR}_k$ (see Sec.\
\ref{ss.before1}) and the last two terms come from the entropy
perturbations.  There are three contributions to the constant mode,
and in the long wavelength limit, the dominant contribution is
almost scale-invariant.

An important point is that while the amplitude of the new terms
in $\mR_k$ sourced by the entropy perturbations depends quite strongly
on the specifics of how $\d\theta \neq 0$, the $k$-dependence of
the two new terms, namely $k^{-3/2+p}$ and $k^{3/2-p}$, is robust so
long as the $k$-dependence of $\de s_k$ does not change while
$\d\theta \neq 0$.

Finally, note that since the prefactors to the terms generated due to
the entropy perturbations highly depend on $t_i$ ---not only due to
the time-dependence of the prefactors in \eqref{mRdot}, but also the
time-dependence in $\de s_k$ itself--- the amplitude of the scale-invariant
mode is also highly dependent on $t_i$.

\subsection{The Bounce}
\label{ss.bounce2}

Using the LQC effective equations, it is possible to evolve the curvature
perturbations through the bounce.  Note that in general relativity,
in the absence of entropy perturbations and for modes that are outside
the Hubble radius, $\mR' = 0$.  However, this is not the case in LQC and
so it is necessary to solve the equations of motion for $\mR$ even in
the absence of entropy perturbations.

Here we will assume that the entropy perturbations do not play a role once
the scale-invariant perturbations have been generated, and then the long
wavelength solution \eqref{R-lqc1} to the LQC effective equation for $\mR$,
\be \label{mR-bounce}
\mR_k = C_1 + C_2 \int^{\b\eta} \f{\dd \b\eta}{z(\b\eta)^2},
\ee
is valid.

Now, since at this point $\vp_2$ has climbed the potential, the scalar fields
no longer mimic an ultrastiff perfect fluid and so the exact dynamics of the
background are not known without specifying the entire form of the potential
$V_2(\vp_2)$ (and also $V_1(\vp_1)$ if that potential is to be modified as
well), and then determining the dynamics of the background metric by using the
LQC effective Friedmann equations.  Because of this, it is not possible to
determine the exact values of $C_1$ and $C_2$ without further input.

Nonetheless, it is possible to determine the $k$-dependence of these terms
in the following manner.  Recall that in Sec.\ \ref{ss.bounce1}, it was shown
how by choosing an appropriate constant of integration, it is possible to
relate $C_1$ only to the constant mode of $\mR$ in the contracting phase
and $C_2$ only with the time-dependent mode of $\mR$; this procedure gives
\be
C_1 \sim k^{-\tf{1}{2}+p} + k^{\tf{3}{2}-p} + k^{-\tf{3}{2}+p}, \qquad
C_2 \sim k^{\tf{1}{2}-p}.
\ee

On the other side of the bounce, the equations of general relativity can be
trusted once more, and the curvature perturbation will contain the two
standard modes, of which the constant mode is the one relevant for structure
formation.

From the solution \eqref{mR-bounce}, it is clear that at least the three terms
in $C_1$ will contribute to the constant mode, and generically the term in
$C_2$ will contribute as well (as it did in Sec.\ \ref{ss.after1}).  Therefore,
after the bounce the constant mode of the curvature perturbation will typically
contain the four following terms (and always the first three terms),
\be
\mR_k \sim k^{-\tf{1}{2}+p} + k^{\tf{3}{2}-p} + k^{-\tf{3}{2}+p}
+ k^{\tf{1}{2}-p},
\ee
and the dominant contribution at long wavelengths is almost scale-invariant, so
\be
\Delta_\mR^2(k) \sim k^{2p}.
\ee
Note that the amplitudes of each term in the constant mode depend on the specifics
of the pre-bounce dynamics.  Furthermore, the background dynamics during the
bounce only affect the amplitude of the last term, which comes from the decaying
mode in the contracting branch.

What can be seen here is that while $\mR' \neq 0$ in LQC ---even for adiabatic
perturbations that are outside the Hubble radius--- the terms already appearing
in the constant mode of $\mR$ are conserved.  Typically, however there are
additional terms that are generated during the bounce.  Note that as the new
terms will have a different dependence on $k$ (except perhaps in some
pathological cases) the new terms cannot affect the amplitudes of the already
existing modes or cancel them out.

In the ekpyrotic universe, the new term added during the bounce has a blue
spectrum, so it is not relevant from an observational point of view and can
be ignored.  Thus, the almost scale-invariant term dominates the scalar
power spectrum in ekpyrotic LQC.

\subsection{More General Potentials}
\label{ss.pot}

Clearly, the discussion so far in this section has only considered a very
specific set of potentials and it may appear that a significant amount of
fine-tuning ---both in the choice of the potentials, and in the initial
conditions so that $\vp_2$ starts climbing its potential at the right
time--- is necessary in order to obtain a scale-invariant spectrum of
curvature perturbations.

However, these choices have been made in order to simplify the calculations
so that the derivation is as clear and transparent as possible.  It is of
course possible to consider different potentials, as well as different
procedures that generate a nonzero $\d\theta$, and some of these choices
will no doubt require less fine-tuning.

Let us clarify which ingredients are necessary in order to obtain
scale-invariant curvature perturbations in ekpyrotic loop quantum
cosmology:
\begin{enumerate}
\item The first ingredient is a contracting phase where the matter degrees
of freedom mimic a perfect fluid with a very large (constant) equation of
state.  If there are at least two matter fields, and the entropic perturbations
begin as quantum vacuum fluctuations, then the entropic perturbations will acquire
a scale-invariant spectrum.
\item Then, the (scale-invariant) entropy perturbations can source new
(scale-invariant) terms in the curvature perturbations.  If the matter fields
are scalars, this occurs if there is a sharp turn in the trajectory of the
scalar fields in their configuration space.
\item Using the LQC effective equations, and assuming that the entropy
perturbations no longer affect the dynamics of the curvature perturbations,
it is possible to evolve the perturbations through the bounce.  It then
follows that the constant mode of the curvature perturbation in the expanding
branch contains an almost scale-invariant term, and it is this term that is
the dominant one for long wavelengths.
\end{enumerate}

There are two further comments that are necessary.  First, the amplitude and
tilt of the almost scale-invariant perturbations strongly depend on the
specifics of (i) the background dynamics of the contracting phase and
(ii) the manner in which $\theta$ changes.  In particular, there are rather
simple examples that give predictions consistent with the observations of
the cosmic microwave background \cite{Lehners:2007ac, Buchbinder:2007ad,
Fonseca:2011qi}.  In any case, there is a lot of freedom in these choices,
and an important point is that these choices can be studied entirely in the
classical limit, as both the tilt and the amplitude remain constant through
the bounce (so long as the entropy perturbations decouple from the curvature
perturbations during the bounce).

The second comment concerns this last caveat: entropy perturbations have not
yet been studied in LQC.  For this reason, in this paper we have assumed that
$\de s$ sources $\mR$ before quantum gravity effects become important, and also
that the two decouple during the bounce.  In order to go beyond these assumptions
it will be necessary to determine the LQC effective equations in the presence
of entropy perturbations.  It seems likely that allowing a coupling between the
curvature and entropy perturbations during the bounce will not affect any
qualitative predictions, but this should be checked.

\section{Conclusions}
\label{s.disc}

In this paper, we have determined the spectrum of curvature perturbations
in ekpyrotic loop quantum cosmology, for both the single scalar field and
the two scalar field model.  While the case of a single field is not viable
as it predicts a blue spectrum, in the two scalar field scenario ---where
entropy perturbations generate additional contributions to the curvature
perturbations--- the resulting power spectrum is indeed scale-invariant and
so is compatible with observations.

In this work, particular potentials for the scalar fields have been chosen
in order to simplify the calculations.  It is of course
possible to consider a larger family of potentials than the ones used here,
at the expense of complicating some of the calculations.  Indeed, it will
be necessary to do just this in order to obtain perturbations with a slight
red tilt to their scale-invariance, as the potentials considered here
give a slight blue tilt.

This is not very problematic however, as some potentials that do give a red tilt
are already known \cite{Lehners:2007ac, Buchbinder:2007ad, Fonseca:2011qi}.  In
addition, as argued in Sec.\ \ref{ss.pot}, any contributions to the constant mode
of the curvature perturbation $\mR$ (the mode that is observationally relevant)
before the bounce will survive the bounce.  There may be additional contributions
to the constant mode of $\mR$, but so long as these are small compared to
$k^{-3/2}$ in the long wavelength limit, the almost scale-invariant term
will dominate.  Furthermore, the amplitude of this mode is entirely determined
by the pre-bounce physics, where the quantum gravity effects are negligible
and the standard cosmological perturbation equations can be used.  Therefore,
in ekpyrotic loop quantum cosmology it is possible to determine the amplitude
of the curvature perturbations and their departure from scale-invariance for
a specific set of potentials purely from the classical equations of motion
(so long as the entropic perturbations source the curvature perturbations
before quantum gravity effects become important) and so the results obtained
for a larger family of potentials given in \cite{Lehners:2007ac, Buchbinder:2007ad,
Fonseca:2011qi} hold in LQC as well.

On the other hand, if the generation of curvature perturbations from the
entropy perturbations occurs at a time when LQC effects are important, then
there may be some additional corrections to the resulting scalar power
spectrum that have not been considered here.  An investigation of this
possibility is left for future work.

\acknowledgments

I would like to thank
Thomas Cailleteau
and
Parampreet Singh
for helpful discussions.
This work was supported in part by
a grant from the John Templeton Foundation.
The opinions expressed in this publication are those of the
author and do not necessarily reflect the views of the John
Templeton Foundation.

%

\end{document}